# Design of a 1×4 CPW Microstrip Antenna Array on PET Substrate for Biomedical Applications

U. Farooq, A. Iftikhar, M. S. Khan, M. F. Shafique, Raed M. Shubair

# ABSTRACT

In this paper, a single layer Coplanar Waveguide-fed microstrip patch antenna array is presented for biomedical applications. The proposed antenna array is realized on a transparent and flexible Polyethylene Terephthalate substrate, has 1×4 radiating elements and measures only 280 × 192 mm2. The antenna array resonates at 2.68 GHz and has a peak-simulated gain of 10 dBi. A prototype is also fabricated, and the conductive patterns are drawn using cost-efficient adhesive copper foils instead of conventional copper or silver nanoparticle ink. The corresponding measured results agree well with the simulated results. The proposed low profile and cost-efficient transmit antenna array has the potential for wearable born-worn applications, including wireless powering of implantable medical devices.

I. INTRODUCTION

Real time health monitoring received extensive attention with the evolution of wireless communications. Wearable sensors and wireless devices, along with some data processing gadgets allow the identification and transmission of human body's vital signs using wireless sensors nodes [1]. The wearable antennas pave the road to the advancements of wireless transmissions systems not only for human body area networks, but they are also employed in Global Positing System (GPS) and Multiple Input Multiple Output (MIMO) antennas (etc.). These antennas are required to be inexpensive, low profile, flexible, and must provide easy integration with the planar circuits to enable practical applications [1, 2]. Among the family of body-worn antennas, the Co-planar Waveguide (CPW) fed antennas exhibit better performance in terms of their bandwidth and radiation characteristics. However, a single antenna cannot meet the desired gain characteristics and hence an antenna array must be utilized for practicality [3].

Microstrip patch antenna arrays are the most common candidate to enhance the antenna gain and are widely applied for wireless power transfer in medical implants. For instance, authors in [4] have realized a 1×2 array antenna to transmit RF energy into the body to remotely power a leadless pacemaker (LP). Similarly, in another study, a CPW fed Ultra-wide band (UWB) antenna is presented for footwear applications to measure the human body changes [5]. A square patch implantable antenna loaded with Complementary Split Ring Resonator (CSRR) is presented in [6]. However, the proposed design is realized on a rigid substrate and cannot be conformed to non-planar surfaces. Moreover, the use of commercially available copper or silver nano-particle ink, for drawing conductive patterns on non-conventional flexible substrates (e.g. fabric, flexible plastic etc.), is also a key challenge for larger scale production of antenna arrays [7]. Therefore, a flexible, cost-efficient antenna array with enhanced gain characteristics is highly desirable, especially for health monitoring and biotelemetry applications. In a fashion similar to [8]-[32], the work in this project is a continuation of the several previous contributions, which have been reported in the literature in various relevant technologies, systems, and their associated applications.

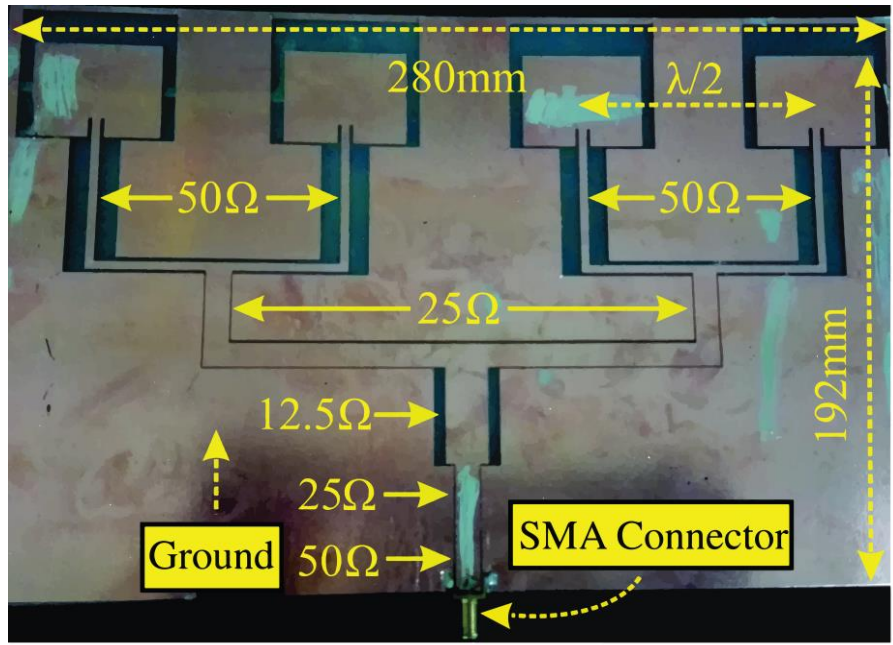

Fig. 1: Layout of the proposed CPW microstrip transmit antenna array.

The objective of this work is to propose an economical and flexible CPW microstrip transmit array antenna for body-worn health monitoring applications and RF pacemakers. The proposed CPW microstrip antenna array, as shown in Fig. 1, is realized on a 1 mm thick transparent and flexible Polyethylene Terephthalate (PET) substrate with a relative permittivity $\epsilon_r$ of 1.8. The conductive patterns are drawn using cost-efficient conductive copper foils. A prototype antenna array measuring only 280×192 mm$^2$ is fabricated and tested. The measured results show good agreement with the simulated results.

## II. CPW-FED ANTENNA ARRAY DESIGN

The proposed CPW fed transmit array antenna is shown in Fig. 1, which comprises a thin copper foil on a single copper layer on a transparent and flexible Polyethylene Terephthalate (PET). A single layer metallization is used to keep the radiation pattern intact with high gain characteristic, which may degrade due to the close proximity to the human body. The design procedure involves the realization of a single CPW microstrip patch antenna and then replicated of single antenna at a center to center distance of λ/4 to form a 1×4 array, fed by a corporate feed network. Fig. 1 shows the four individual patch antennas are fed by a 50Ω microstrip inset feed, whose length and width are optimized to achieve the maximum impedance match between the feed and the radiating patch. The two 50 Ω feed lines in parallel to each other result in a 25 Ω feed line whose length and width is calculated by experimentally verified formulas. Afterwards, the two 25Ω feed lines produce 12.5 Ω net impedance that is transformed to 50 Ω feed at source input using a quarter-wave transformer. The proposed CPW microstrip transmit array antenna is simulated in a full-wave 3D Finite Element Method (FEM) based electromagnetic simulator.

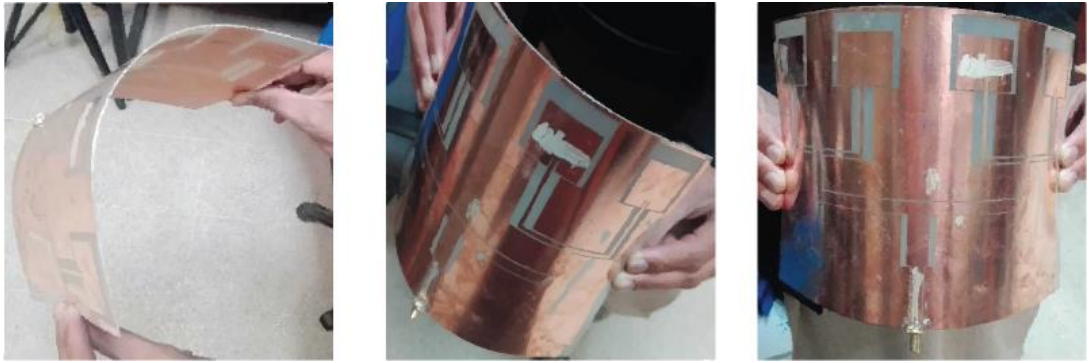

Fig. 2: Pictures of the fabricated prototype showing conformability of the proposed transmit antenna array.

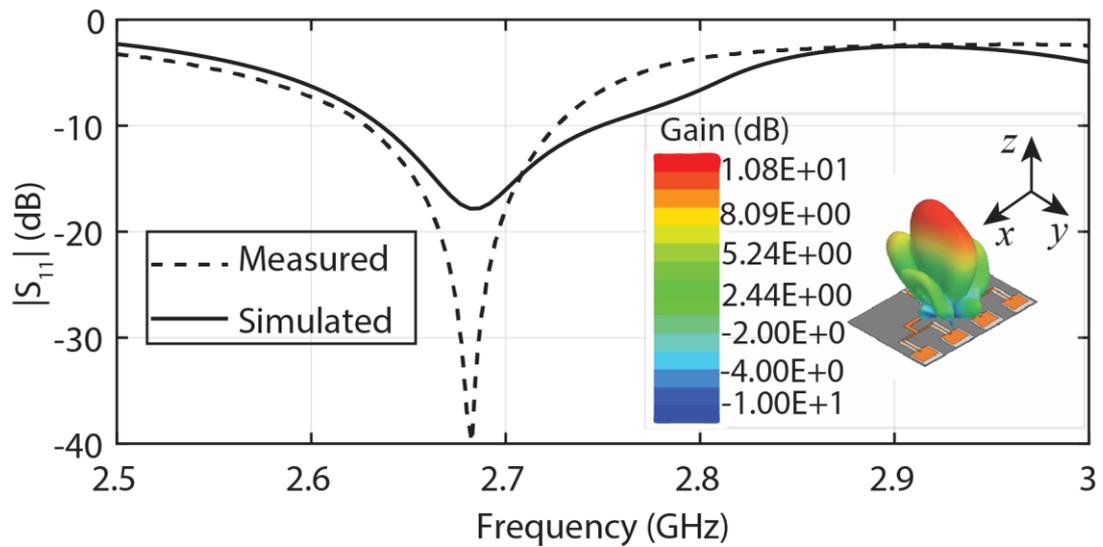

Fig. 3: Comparison of simulated and measured $|S_{11}|$ (dB) of proposed CPW microstrip antenna transmit array.

## III. PROTOTYPING AND RESULTS

The prototype of proposed CPW-fed antenna array is fabricated and tested using a fully calibrated vector network analyzer (Agilent N5242A). Initially, the substrate is washed with soapy water to remove impurities prior to lamination of adhesive copper foils. Next, the adhesive copper foils were laminated through heat laminator on flexible PET substrate. Eventually, the standard chemical etching process was used to draw conductive patterns. Pictures of the fabricated prototype showing conformability of the proposed CPW microstrip antenna array are shown in Fig. 4. Comparison of the simulated and measured input reflection coefficients ($|S_{11}|$ in dB) shows a good agreement, as shown in Fig. 3. It shows that the proposed array antenna resonates at 2.68 GHz with overall bandwidth (-10 dB) of more than 100 MHz. Fig. 3 (inset figure) also shows the simulated gain of array antenna, which is approximately 10 dBi. To study the directional properties of the proposed antenna array, the E-plane and H-plane radiation pattern is investigated and shown in Fig. 4. The small deviations in measured results are attributed to fabrication tolerances and measurement imperfections.

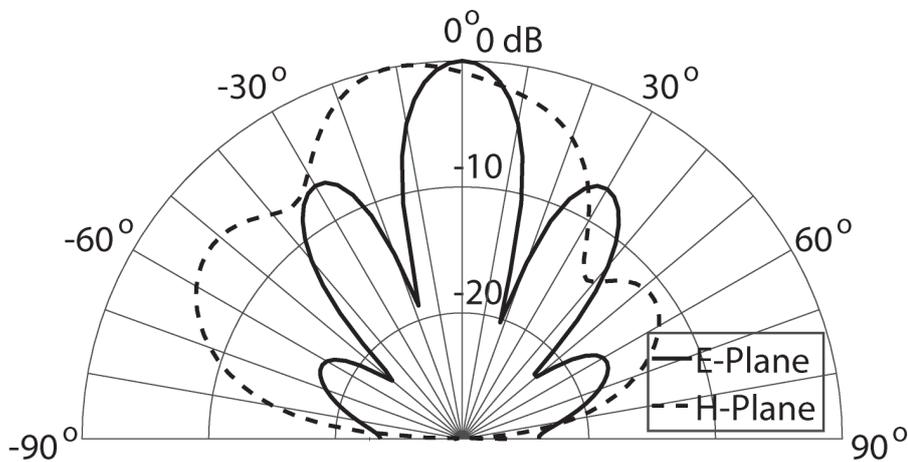

Fig. 4: Simulated E- and H-plane radiation patterns of the proposed CPW microstrip antenna transmit array at 2.68 GHz.

## IV. CONCLUSION

A CPW-fed 1×4 microstrip antenna array is presented in this paper. The proposed array antenna is realized on a flexible PET substrate and conductive patterns are drawn using adhesive copper foils instead of conventional ink (silver/copper nanoparticle ink). The prototype antenna array showed good impedance matching when tested in conformal conditions and hence demonstrate a good potential for wearable applications. Future study will focus on further optimization of the proposed design for Industrial, Scientific, and Medical (ISM) band applications. Additionally, the proposed transmit antenna array will be investigated for applications in wireless powering of implantable medical devices as well as for biomedical telemetry. In a fashion similar to [8]-[32], the work in this project is a continuation of the several previous contributions, which have been reported in the literature in various relevant technologies, systems, and their associated applications.